\documentclass[aps,prb,twocolumn,showkeys,groupedaddress]{revtex4-1}
\newcommand{\squeezeup}{\vspace{0cm}}
\newcommand{\chgd}[1]{{#1}}
\newcommand{\chgdtwo}[1]{{#1}}

\usepackage{graphicx}
\usepackage{xcolor}
\graphicspath {{Figures/}}
\usepackage{epstopdf}

\begin{document}


\title{On the Lorenz number of multi-band materials}


\author{Mischa Thesberg}
\affiliation{Institute for Microelectronics, TU Wien, Austria}

\author{Hans Kosina}
\affiliation{Institute for Microelectronics, TU Wien, Austria}

\author{Neophytos Neophytou}
\affiliation{School of Engineering, University of Warwick, Coventry, UK}


\date{\today}

\begin{abstract}

There are many exotic scenarios where the Lorenz number of the Wiedemann-Franz law is known to deviate from expected values.  However, in conventional semiconductor systems, it is assumed to vary between the values of ${\sim}1.49\times 10^{-8}$ W $\Omega$ K$^{-2}$ for non-degenerate semiconductors and ${\sim}2.45 \times 10^{-8}$ W $\Omega$ K$^{-2}$ for degenerate semiconductors or metals. Knowledge of the Lorenz number is important in many situations, such as in the design of thermoelectric materials and in the experimental determination of the lattice thermal conductivity.  Here we show that, even in the simple case of two and three band semiconductors, it is possible to obtain substantial deviations of a factor of two (or in the case of a bipolar system with a Fermi level near the midgap, even orders of magnitude) from expectation.  In addition to identifying the sources of deviation in unipolar and bipolar two-band systems, a number of analytical expressions useful for quantifying the size of the effect are derived.  As representative case-studies, a three-band model of the materials of lead telluride (PbTe) and tin sellenide (SnSe), which are important thermoelectric materials, is also developed and the size of possible Lorenz number variations in these materials explored.  Thus, the consequence of multi-band effects on the Lorenz number of real systems is demonstrated.
\end{abstract}


\maketitle

The Wiedemann-Franz law connects the electronic part of the thermal conductivity ($\kappa_e$) to the electronic conductivity $\sigma$ through the relation $\kappa_e = L \sigma T$, where $L$ is the Lorenz number. For metals and degenerate semiconductors, $L$ reaches the Sommerfeld value, $L_0=\pi^2/3 (k_B/q)^2=2.45 \times 10^{-8}$ W $\Omega$ K$^{-2}$ where $k_B$ and $q$ are the Boltzmann constant and charge of an electron, respectively. This value drops to $L_0= 2 (k_B/q)^2 =1.49\times 10^{-8}$ W $\Omega$ K$^{-2}$ for non-degenerate, single parabolic band materials and acoustic scattering conditions.\cite{Price57, Kim15} 

The Lorenz number plays an important role in the experimental determination of the phonon, or lattice part of the thermal conductivity ($\kappa_L$) from thermal conductivity measurements , which is done by computing and subtracting the electronic part ($\kappa_e$) from the experimentally measured value of the total thermal conductivity ($\kappa_{tot}$). Thus, deviations in its value result in incorrect determination of the relative contributions of charge and phonons to heat flow in real materials.

An example of a case where this issue plays an important role is in thermoelectric (TE) materials engineered to harvest renewable energy from waste heat. Thermoelectric efficiency is optimized by lowering the thermal conductivity as low as possible while keeping electrical conductivity high. Through this lense, one can interpret the Lorenz number as a quantification of the reality that one cannot raise electrical conductivity without proportionately increasing the electronic thermal conductivity as well. Thus, an understanding of deviations from expected Lorenz values allows: i) a direct means of enhancing TE performance by finding cases of lower Lorenz number,\cite{Flage11} and ii) a more accurate estimate of the lattice thermal conductivity. The second point is especially important as there have been extensive efforts\cite{Snyder08,Vineis10,Biswas12,Zhang13,Kim15,Xu15} towards improving TE performance by nano-structuring materials to lower their thermal conductivity (in addition to potentially enhancing their electrical performance)\cite{Neophytou2011,Thesberg2015,Thesberg2016} and several claims of phonon thermal conductivities below the amorphous limit of 1-2 W/mK have been reported. These next generation TE materials exhibit a degree of complexity in their electronic structure and dominant scattering mechanisms. Therefore, there is no reason to expect that the Sommerfeld value holds for them, nor that the Lorenz number's value can only fall intermediately between the two limiting values.\cite{Flage11}

In fact, deviations in the Lorenz number are nothing new and occur in various cases: confined dimensions\cite{Ou08, Tripathi10} such as nano-wires,\cite{Volklein09,Cheng15,Li11} quasi-1D systems,\cite{Casian10,Wakeham11,Vavilov05} and effective 0-D systems such as quantum dots,\cite{Lopez13,Krawiec06,Trocha12}, single molecule\cite{Wang10,Silva12,Wierzbicki11} and single atom\cite{Kubala08} systems; under conditions of quantum criticality;\cite{Tanatar07, Kim09, Machida13, Khodel08, Maestro08} in superconductors;\cite{Graf96,Houghton02, Durst00, Sologubenko02,Bel04,Lee04,Lee03,Li11} in superlattices and granular metals;\cite{Wang13, Beloborodov05, Zhang06,Bian07, Tripathi06} and in the presence of disorder.\cite{Sun09,Garg09}  However, here we consider a far more common, but yet to be explored, situation - the case of multi-band semiconductors, which are the most relevant case for next generation thermoelectric materials with complex bandstructures.\cite{Zhao14,Zhao16,Guo15,Pei11} 

In this work, we explore the issue of variations in the Lorenz number in materials that contain more than one band, different types of scattering, and different band effective masses. Such materials with complex bandstructures, such as SnSe,\cite{Zhao14,Zhao16} SnS,\cite{Guo15} PbTe,\cite{Pei11} etc, are currently receiving large attention for TE applications.\cite{Vineis10,Snyder08} \chgd{We employ the Boltzmann transport method, expressed in the Landauer form} to examine a number of different cases: i) a system of two conduction bands in the absence of inter-band scattering, ii) a bipolar system of one conduction band and one valence band, which was shown in Ref.~\cite{Jeong10} to accurately model BiTe, iii) a three-band model (one conduction and two valence bands) of the common thermoelectric material lead telluride (PbTe), without inter-band scattering, iv) a case of a two conduction band system, in the presence of inter-band scattering, and v) a three-band model (one conduction and two valence bands) of in tin selenide (SnSe), with inter-band scattering. In all cases we show that significant variations are observed from the interval between 1.49 (non-degenerate limit) to 2.44 (degenerate limit) W $\Omega$ K$^{-2}$ with some being as high as 100\% deviation in unipolar materials. \chgd{In the case of bipolar materials, it is well understood that the Lorenz number deviates from either limit. Here we show that this deviation can reach orders of magnitude when the Fermi level is close to the mid-gap and also provide simple analytical formulae to quantify these deviations.} 

\chgdtwo{Crucially, we show that the simplified formula:
\begin{equation}\label{Eq:Wrong_L_Formula}
L \approx \frac{\sum_i L_i G_i}{\sum_i G_i},
\end{equation}
\noindent where $L_i$ and $G_i$ are the Lorenz number and conductance of the $i$th bands respectively, which has seen some use in the literature\cite{Pei11,Zhang13,Xu15} as an approximation of the Lorenz number, omits a crucial term that couples the multiple bands, even in the absence of inter-band scattering.  This missing term is significant - especially in bipolar systems where the well known bipolar effect occurs.}

Our results will allow for better estimates and understanding of the lattice thermal conductivity, especially in materials relevant to thermoelectricity such as SnSe, SnS, PbTe and BiTe\cite{Vineis10, Guo15} but also how the Lorenz number behaves in general in materials with complex band-structures. 

\section{Methods}

\subsection{The Landauer Formalism}

\chgd{In this work we exclusively consider parabolic effective mass systems  in the linear response regime.  Transport is described by the Boltzmann Transport Equation expressed in the Landauer form in terms of an effective transmission and number of transmitting channels.\chgdtwo{\cite{Datta97,Lundstrom_Book13,Jeong10,Jeong11,Maassen13}} As we explore below, the Landauer method can be formulated such that it maps to the BTE even in the diffusive regime, so alternatively we can call such an approach simply `the Landauer formalism within the diffusive regime', and that terminology will be used regularly.} Within this formalism it is possible to define analytic expressions for the important thermoelectric parameters in terms of integrals of the form:
\begin{equation}
\label{Eq:I_integral_definition}
I_j = \int_{- \infty}^{\infty} \eta_F^j \overline{T}(\eta) \left( - \frac{\partial f_0}{\partial \eta} \right) d\eta
\end{equation}
\noindent where $\eta$ and $\eta_F$ are the reduced band energy and reduced Fermi level respectively:
\begin{equation} 
\eta = \frac{E-E_b}{k_B T}, \qquad  \eta_F = \frac{E_F-E_b}{k_B T}
\end{equation} 
with $E_b$ being the band energy (i.e. $E_{C_1}$, $E_V$, etc.), be it conduction ($E_C$) or valence ($E_V$), $k_B$ being the Boltzmann constant and $T$ being the temperature. $f_0$ represents the Fermi-Dirac distribution and $\overline{T}(E)$ is the \emph{effective} transmission. Within the \chgd{Landauer formalism in the diffusive limit} the effective transmission is given by:
\begin{equation}
\label{Eq:Effective_Transmission}
\overline{T}(E) = T(E) M(E)
\end{equation}
\noindent where $T(E)$ is the transmission and $M(E)$ is the density of modes,\cite{Lundstrom_Book13} which in three dimensions for parabolic bands \chgd{(excluding spin degeneracy)} is:
\begin{equation}
\label{Eq:3D_DOM}
M_{3D}(E) = A \frac{m^*_{DOM}}{2 \pi \hbar}(E-E_b)
\end{equation}
\noindent with $m^*_{DOM}$ being the \emph{density-of-modes} effective mass,\cite{Lundstrom_Book13} $A$ being the cross-section area of transport and $\hbar$ being the reduced Planck constant. In this work we are only concerned with conductances rather than conductivities and thus size dependences introduced by areas, such as $A$, and length, $\ell$, are removed.

Although the Landauer formalism allows one to treat both diffusive and ballistic systems, here we focus on the diffusive regime, which is representative of room temperature transport. In that case, the transmission function $T(E)$ can be assumed to be:
\begin{equation}
\label{Eq:Transmission}
T(E) = \frac{\lambda(E)}{\ell}
\end{equation}
\noindent where $\lambda(E)$ is the \emph{mean-free-path for back-scattering}\cite{Jeong10,Lundstrom_Book13} and $\ell$ is the system length, which we again remove. The relationship between the mean-free-path for back-scattering and the more conventional scattering time is given by the simple expression (in three dimensions, \chgd{under the assumption of isotropic energy bands}):
\begin{equation}
\label{Eq:MFP_wrt_tau}
\lambda(E) = \frac{4}{3} v(E) \tau(E)
\end{equation}
\noindent where $v(E)$ is the velocity of carriers and $\tau(E)$ is the scattering time.

\chgd{We emphasize that the Landauer formalism, with a transmission function defined by a semi-classical power law relation used here, ends up being mathematically equivalent, though conceptually distinct, from the more common Boltzmann transport approach.} Specifically, the effective transmission in the Landauer approach is related to the transport distribution function of Boltzmann transport through the simple expression:
\begin{equation}
\Xi(E) = \frac{\ell^2}{h} \overline{T}(E) = \frac{\ell}{h} \lambda(E) M(E)
\end{equation}
\noindent in this limit.  \chgd{However, as the Landauer approach allows one to capture ballistic transport through a different choice of the transmission and as it is framed not in terms of more classical notions of carrier velocity, but rather quantum modes, we find it to be a more natural language for the field of thermoelectrics in general, which is lately dominated by considerations of low-dimensional- and nano-structures.}
 
\subsection{Scattering}

In this work we assume that $\lambda(E)$ has the simple, commonly employed, form:
\begin{equation}
\label{Eq:Lambda_temp_form}
\lambda(E) = \lambda_0'(T) \eta^r
\end{equation}
\noindent where $r$ is an integer exponent ($r=0$ for acoustic phonons in three-dimensions) and $\lambda_0'$ is a constant.  \chgd{In this work, for conceptual simplicity, all bands are assumed to be parabolic.} This is equivalent to, although easier to justify than,\cite{Jeong10} the common assumption in Boltzmann transport theory that the scattering time is of the form:
\[\tau = \tau_0 \eta^s. \]

For the case of acoustic phonon scattering, Boltzmann theory dicatates that $\tau$ is given by:
\begin{eqnarray}
\tau_{AP,b}(E) =  \frac{\beta_{AP}}{k_B T (m^*_b)^{3/2} D_{AP,b}^2} \frac{1}{\sqrt{E-E_b}}, 
\end{eqnarray}
\noindent where $D_{AP,b}$ is the deformation potential of the acoustic phonons of band $b$, $m_b^*$ is the \emph{density-of-states} effective mass of band $b$ and $\beta_{AP}$, given by:
\begin{equation}
\label{Eq:Beta_AP}
\beta_{AP} = \frac{\pi \hbar^4 \rho c_s^2}{\sqrt{2}},
\end{equation}
\noindent which encapsulates all material properties and constants that do not depend on the specific band: $\rho$ the mass density and $c_{s}$, the sound velocity.  Using Eq.~\ref{Eq:MFP_wrt_tau} one can re-write this as a mean-free-path to get:
\begin{eqnarray}
\lambda_{AP,b}(E) &=& \frac{4}{3} \frac{\sqrt{2}\beta_{AP}}{k_B T (m^*_b)^2 D_{AP,b}^2}\\ 
 &=& \frac{\lambda_{AP}}{k_B T (m^*_b)^2 D_{AP,b}^2} \eta^0,
\end{eqnarray}
\noindent where $\lambda_{AP}$ collects all the constants associated with the \chgd{electron-phonon scattering} behaviour of the material, $\lambda_{AP} = \sqrt{32}\beta_{AP}/3$ and can either be treated as a tunable parameter, to match experiment, or explicitly calculated based on knowledge of the constants in Eq.~\ref{Eq:Beta_AP}.

Generalizing these results to the case of any scattering mechanism defined by a power law of exponent $r$, we define the mean-free-path for backscattering in a band $b$ as:
\begin{equation}\label{Eq:MFP_Real}
\lambda_{b}(E) = \frac{\lambda_0}{(k_B T)^{1-r} (m^*_b)^2 D_{AP,b}^2} \eta^r,
\end{equation}
\noindent where $\lambda_0$ is both material and scattering mechanism dependent and all quantities with a subscript $b$ are band dependent. \chgd{In this way the diffusive BTE is transformed into the diffusive Landauer form, and by assigning a mean-free-path a more physical understanding is provided in comparison to relaxation times.} Throughout this work, unless a particular material is being considered, $\lambda_0/D_{AP,b}^2$ is arbitrarily chosen for all bands such that the resulting mean-free-path for back-scattering at $300$ K of a band with effective mass of $m_0$ is $20$ nanometers (i.e. $\lambda_0/ D_{AP,b}^2 = 20$ nm$ \times k_B (300 $K$) m_0^2$).  Such a value is consistent with many common semiconductors, such as silicon.

This definition of the mean-free path for back-scattering is important. Although it is common when applying the Landauer formalism to simply quote a value for $\lambda(E)$ directly, looking at Eq.~\ref{Eq:MFP_Real} it is clear that the mean-free-path of the carriers associated with a given band, scales as the effective mass is changed. Thus, in working in multi-band systems with different band effective masses, in order to be more accurate, one must consider how the mean-free-path of each of the bands scales accordingly.

\subsection{Transport Coefficients}

Using the integrals $I_j$ (Eq.~\ref{Eq:I_integral_definition}), the various electronic transport coefficients can be defined as: 
\begin{eqnarray} 
\label{Eq:G_definition}
G = (2 q^2 / h) I_0\qquad &[1/\Omega],&\\
\label{Eq:SG_definition}
\overline{SG} = -(k_B/q)I_1\qquad &[\textrm{V}\; \Omega^{-1} \textrm{K}^{-1}],& \\
\label{Eq:S_definition}
S = \overline{SG}/G \qquad &\textrm{[V/K]},&\\
\label{Eq:kappa_0_definition}
\kappa_0 = (T 2 k_B^2/h)I_2 \qquad &[\textrm{W/K}],&\\
\label{Eq:kappa_e_definition}
\kappa_e = \kappa_0 - TS^2 G \qquad &[\textrm{W/K}],&\\
\label{Eq:L_definition}
L = \kappa_e/(T G)\qquad &[\textrm{W}\; \Omega \; \textrm{K}^{-2}]&
\end{eqnarray}

We refer to these quantities as: the electrical conductance ($G$), the Soret coefficient for electro-thermal diffusion ($\overline{SG}$), the Seebeck coefficient ($S$), the short-circuit electronic thermal conductance ($\kappa_0$), the electronic thermal conductance for zero electric current ($\kappa_e$) and the Lorenz number, respectively.

In the absence of inter-band scattering a multi-band system can be modeled within the Landauer formalism by simply treating each band's density-of-modes separately (i.e. $M(E) = M_1(E) + M_2(E) + \ldots$). In this way, we find for a two-band system that:
\begin{eqnarray}
G_{tot} &=& G_1 +G_2 \\
\label{Eq:S_two_band}
S_{tot}  &=& \frac{\overline{SG}_1 + \overline{SG}_2}{G_{tot}} = \frac{S_1 G_1 + S_2 G_2}{G_{tot}},
\end{eqnarray}
\noindent where the second line results from the fact that $\overline{SG}_i = \overline{SG}_i (G_i/G_i) = S_i G_i$. Looking at these two expressions gives the impression that in a system where carriers in one band do not scatter into the other, there is no coupling between bands and each thermodynamic quantity can be treated as a sum weighted by each band's relative contribution to the conductance. \chgd{In this way one might assume the multi-band Lorenz number takes the plausible form:}
\begin{equation}
L \approx \frac{L_1 G_1 + L_2 G_2}{G_1+G_2}.
\end{equation}

However, these first two quantities create an erroneous impression as can be seen when one attempts to calculate $\kappa_e$ for a two band system:
\begin{eqnarray}
\kappa_{e,tot} = \kappa_{0,1} + \kappa_{0,2}  - T S_{tot}^2 G_{tot}
\end{eqnarray}

The crucial aspect here is the term $S_{tot}^2$, which, by looking at Eq.~\ref{Eq:S_two_band}, we can clearly see that it must contain terms $\propto S_1 S_2$. As a result of this, the electronic thermal conductance ($\kappa_e$) and thus the Lorenz number ($L$) \emph{cannot} be treated according to Eq.~\ref{Eq:Wrong_L_Formula}; even if a multi-band system has no explicit inter-band scattering, bands are still coupled to one another.

\section{Results and Discussion}
Below we explore the quantitative and qualitative deviations from expected values of the Lorenz number in a number of different cases of multi-band systems with substantial deviations and unexpected behaviour found in all cases. In addition, the size of these deviations will be quantified in the real thermoelectric materials of lead telluride (PbTe) and tin selenide (SnSe). We will also highlight the quantitative discrepancies which result from the application of an expression like Eq.~\ref{Eq:Wrong_L_Formula}, \chgd{which completely ignores multi-band effects,} versus a more correct treatment of $\kappa_e$ and describe the error this will cause in interpreting experimental results.

In Section \ref{Non_Interacting_Bands} we consider the case of multiple bands in the absence of inter-band scattering (i.e. only intra-band acoustic phonon scattering), as well as a case-study of the TE material PbTe. Deviations in the presence of inter-band scattering is the topic of Section \ref{Interacting_Bands} as well as the real material SnSe.

\subsection{\label{Non_Interacting_Bands}Two Bands - Intra-band Scattering Only} 

In order to understand the possible deviations of the Lorenz number that can occur in systems, we initially consider only intra-band scattering, with the two separate cases: i) that of a two-conduction band system and ii) that of a bipolar conducting (i.e. one conduction band, one valence band) system. We will find that in both cases there are important deviations from the expected non-degenerate values. \chgd{In the case of bipolar materials this is expected but for both cases we derive analytical expressions for the deviations based on simple bandstructure features.} Finally, we will consider the case of a real material. For this we pick $p$-type lead telluride (PbTe), which is an important TE material and whose bandstructure we model with two valence bands and a conduction band.  We compute the quantitative effect these deviations can have on predictions of the Lorenz number and, by extension, estimates of the lattice thermal conductivity ($\kappa_L$). We also show that the Lorenz number can deviate as much as an order of magnitude.  As a result, comparatively large overestimations of $\kappa_L$ are possible.

\subsubsection{Two bands of the same type: The Lorenz number in the non-degenerate limit}
\begin{figure}[t]
\squeezeup \squeezeup
\includegraphics[width=0.45\textwidth]{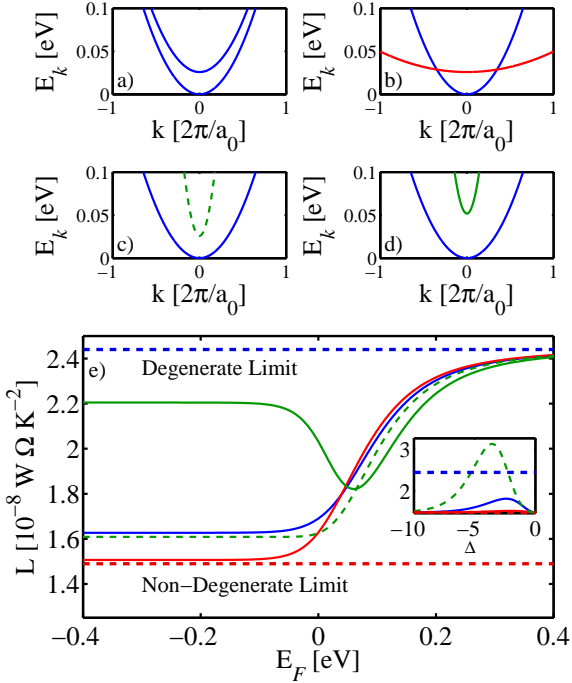}
\squeezeup
\caption{\label{Fig:Non_Interacting_2CB} Lorenz number vs. Fermi level for two conduction bands, with only intra-band scattering. Panels a),b),c) and d) show the energy dispersion of the two-band system with a second band of differing effective mass. In all panels the lowest conduction band has an effective mass of $m_0$ and the second band an effective mass of $m_0$, $10m_0$, $0.1m_0$, $0.1m_0$ for panels a), b), c) and d) respectively, and the energy separation between the first and second bands is $k_BT$ for all panels except for d), where it is $2k_BT$. Panel e) shows the Lorenz number as a function of Fermi level with each curve corresponding to the set-up shown in panels a)-d) (i.e. the blue solid line is the system shown in panel a), red solid line is b), dashed green is c) and solid green is d). Also shown are the degenerate (dashed blue) and non-degenerate (dashed red) limiting values of the Lorenz factor. \chgd{The inset shows the Lorenz number versus the reduced band offset, $\Delta = (E_{C_1} - E_{C_2})/k_B T$, as described by Eq.~\ref{Eq:Lorenz_2band_non_degenerate}.}} 
\end{figure}
As a first example of the effect of multiple bands on the Lorenz number, we consider a system of two conduction bands of differing band edge energies, with intra-band scattering but no inter-band scattering between them, and compute the deviations that occur from the expected non-degenerate Lorenz value of $1.49 \times 10^{-8}$ W $\Omega$ K$^{-2}$. The Lorenz number in such a system can be seen in Figure~\ref{Fig:Non_Interacting_2CB}e, which shows its value as a function of the Fermi level for different permutations of the upper band's effective mass and band separation. The lower band's effective mass is fixed at $m^*_1=m_0$. For Fig.~\ref{Fig:Non_Interacting_2CB}a-c the two bands are separated by an energy of $k_B T \sim 26$ meV ($T = 300$ K) with Fig.~\ref{Fig:Non_Interacting_2CB}a showing the case where both bands have the same mass, Fig.~\ref{Fig:Non_Interacting_2CB}b showing the case where the upper band is substantially heavier (by a factor of 10x) and Fig.~\ref{Fig:Non_Interacting_2CB}c showing the case where it is substantially lighter (by a factor of 10x). 

The expected non-degenerate limit in Fig.~\ref{Fig:Non_Interacting_2CB}e is shown by the dark red dashed line. It is clear in the case of a lighter or similar upper band (the green or blue curves, respectively) that the Lorenz number can saturate in the non-degenerate limit at a higher value than expected. The size of this increase appears to grow as the band separation increases, which can be seen in Fig.~\ref{Fig:Non_Interacting_2CB}d where $m_2^* = 0.1m_1^*$ and the energy separation is $2k_B T$ (corresponding to the solid green line in Fig.~\ref{Fig:Non_Interacting_2CB}e). In that case the difference from the expected non-degenerate limit of $1.49 \times 10^{-8}$ W $\Omega$ K$^{-2}$ is as large as 50\%, with the non-degenerate saturation value nearing the degenerate value of $2.44 \times 10^{-8}$ W $\Omega$ K$^{-2}$ instead. Note that this can provide substantial deviations in the extraction of the lattice thermal conductance, $\kappa_L$, from experimental data of the total thermal conductance ($\kappa_{tot} = \kappa_e + \kappa_L$). As $L$ is larger than thought to be, one could, for example, erroneously assume that $\kappa_L$ is not yet at the amorphous limit and can still be further lowered, when in reality it is already there.

These deviations can be understood using a very simple model of two non-interacting parabolic bands. Such a model is all that is necessary to qualitatively demonstrate the effect and the analytical predictions it makes can be expected to be accurate in the limit of weak inter-band scattering \chgd{which is discussed further on.}

First we introduce the simplifying constants:
\begin{equation}
\alpha_b = m_b^* \lambda_b(E)  k_B T =  \frac{\lambda_{0}}{m^*_b D_{AP,b}} , \textrm{   and   } \gamma = \frac{2 q^2}{h^2},
\end{equation}
\noindent where the values are the same as in Eq.~\ref{Eq:MFP_Real}. These constants effectively divide all material and temperature dependent properties into the constant $\alpha_b$. Using these constants, the conductance of a single parabolic band $b$ with reduced Fermi level $\eta_F = (E_F - E_b)/k_B T$ is given by: 
\begin{equation}
G_b = \gamma \Gamma(r+2)\alpha_b \mathcal{F}_r(\eta_{F_b})
\end{equation}
\noindent where $\Gamma$ is the gamma function, $r$ is the scattering exponent of Eq.~\ref{Eq:Lambda_temp_form} and $\mathcal{F}_r$ is the Fermi-Dirac integral:
\begin{equation}
\label{Eq:FD_Integral}
\mathcal{F}_j(\eta_F) = \frac{1}{\Gamma(j+1)} \int_0^{\infty} \frac{\eta^j}{\exp(\eta-\eta_F) + 1} d\eta.
\end{equation}

  In the case of two conduction (or two valence) bands, with a reduced  conduction band energy offset of $\Delta = (E_{C_1}-E_{C_2})/k_B T$, we can say that $\eta_{F_1} = \eta_F$ and $\eta_{F_2} = \eta_F + \Delta$.

In the non-degenerate limit where $\eta_{F} \ll 0$, $\mathcal{F}_r (\eta_{F_i}) \rightarrow \exp(\eta_{F_i})$. Under this approximation the expression for the conductance simplifies to:
\begin{equation}
G_b \approx \gamma \Gamma(r+2)\alpha_b e^{\eta_{F_b}}.
\end{equation}
For independent bands, the values of the integrals $I_j$ of Eq.~\ref{Eq:I_integral_definition} can simply be added to one another to obtain the full integral for the two-band system. Thus, the total conductance for the two-band system is:
\begin{equation}
\label{Eq:G_both_bands}
G = G_1 + G_2 = \gamma \Gamma(r+2) e^{\eta_F} \psi
\end{equation}
\noindent where
\begin{equation}
\psi = \alpha_1 + \alpha_2 e^{\Delta}.
\end{equation}
From here the final expressions, whose derivation is given in the Appendix, can be obtained:
\begin{eqnarray}
\label{Eq:S_both_bands}
S &\approx & -  \frac{k_B}{q}\left( (r+2) - \eta_F - \frac{\Delta \alpha_2 e^{\Delta}}{\psi} \right) \\
\label{Eq:kappa_both_bands}
\kappa_e &\approx & T \left(\frac{k_B}{q} \right)^2 \gamma \Gamma(r+2)e^{\eta_F} \psi \\* \nonumber &\times & \left[ (r+2)  +\frac{\Delta \alpha_2 e^{\Delta}}{\psi} \left( \Delta - \frac{\Delta \alpha_2 e^{\Delta}}{\psi}\right) \right].
\end{eqnarray}
Finally, by dividing Eq.~\ref{Eq:kappa_both_bands} by Eq.~\ref{Eq:G_both_bands} (times the temperature $T$) we get, by Eq.~\ref{Eq:L_definition}, the final expression for the Lorenz number:
\begin{equation}
\label{Eq:Lorenz_2band_non_degenerate}
L \approx L_0 + \left(\frac{k_B}{q} \right)^2 \left( \frac{\Delta}{1 + \frac{\alpha_1}{\alpha_2}e^{-\Delta} } \right)^2 \frac{\alpha_1}{\alpha_2} e^{-\Delta},
\end{equation}
\noindent where $L_0$ is the typical non-degenerate value of $(r+2)(k_B/q)^2$, which is $ \sim 1.44 \times 10^{-8}$ W $\Omega$ K$^{-2}$ for the case of acoustic phonons in three dimensions where $r=0$. \chgd{Note that, as the non-degenerate limit was invoked in the derivation of these expression, the exponent $r$, which relates the energy to the scattering strength, does not matter.  Thus, this result is considered general for any value of $r$, be it acoustic phonons or weakly or strongly screened impurities.}
\begin{figure}[t]
\squeezeup \squeezeup
\includegraphics[width=0.45\textwidth]{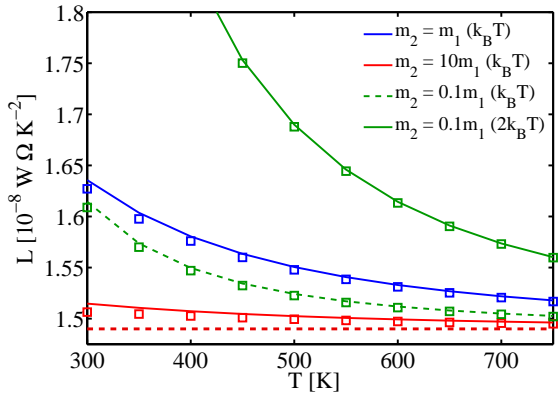}
\squeezeup \squeezeup
\caption{\label{Fig:2CB_L_Analytical} Lorenz number versus temperature for the band structures in Fig.~\ref{Fig:Non_Interacting_2CB}a-\ref{Fig:Non_Interacting_2CB}d. Squares represent numerical results and lines represent the results from the analytical expression in Eq.~\ref{Eq:Lorenz_2band_non_degenerate}. Color and line conventions match those of Fig.~\ref{Fig:Non_Interacting_2CB} with solid and dashed green representing an upper band effective mass $0.1\times$ that of the lower, with an energy separation of $k_B T$ for the dashed line and $2 k_B T$ for the solid line. Blue and red solid lines represent upper band effective masses of $1 \times$ and $10 \times$ that of the lower, respectively, (both at $k_B T$ energy separation).}
\end{figure}
Looking at Eq.~\ref{Eq:Lorenz_2band_non_degenerate} we can see that the existence of a second band, regardless of an assumption of explicit inter-band scattering or not (as in this example), produces a deviation from the expected non-degenerate value of the Lorenz number.  The accuracy of Eq.~\ref{Eq:Lorenz_2band_non_degenerate} is demonstrated in Figure~\ref{Fig:2CB_L_Analytical}, where that expression (solid lines) is plotted alongside those from numerical calculation (squares). The band configurations and line color schemes we employ are the same as with that of Fig.~\ref{Fig:Non_Interacting_2CB}. The analytical expression matches numerics with great precision.

The fact that this occurs can be understood straightforwardly.  Given that in the non-degenerate limit, where the difference between Fermi-Dirac integrals of different type (i.e. $\mathcal{F}_r$ for different $r$) disappears, we still find that in a single band system the ratio of $\kappa_e$ and $TG$ is a fixed constant.  In a two-band system you have an additional term of $- T S_1 S_2 G_{tot}$ arising from the $- T S_{tot}^2 G_{tot}$ term in Eq.~\ref{Eq:kappa_e_definition}.  Given that the term dependent on $\Delta$ in Eq.~\ref{Eq:S_both_bands} is negative, and only occurs in $S_2$ (if we assume the 2nd band is the offset band), then the $\Delta$ dependent portion of the $- T S_1 S_2 G_{tot}$ term will contribute positively to $\kappa_e$ and thus increase the Lorenz number from the single band case.  Thus, this $\Delta$ offset increases the Lorenz number from the expected non-degenerate limit.  Intuitively, this reflects the fact that $\kappa_e$ weights higher energy contributions more compared to the conductance and thus having an increased density-of-states at higher energies (i.e. a second band) improves $\kappa_e$ more than $G$. \chgd{The inset of Fig.~\ref{Fig:Non_Interacting_2CB}e shows how the value of the Lorenz number changes as a function of the reduced band separation, $\Delta$.  It can be seen for the case of a higher band whose effectives mass is ten times smaller than the lower band, that the enhancement above expectation can be over $\sim 100\%$ for a maximum value of $\sim 4 k_B T$. For larger band offsets, the second band is out of the relevant transport energy range and its influence disappears. Note that the effect of the second light band on the Lorenz number is significant, despite the fact that its occupation is minor (due to its low mass and its higher energy) compared to the occupation of the lower band. The light mass allows high velocities which make the upper band similarly conductive to the lower band. Ultimately, however, the effect on the Lorenz number originates from the coupled $S_1S_2$ term as explained above.}

\begin{figure}[t]
\squeezeup
\includegraphics[width=0.45\textwidth]{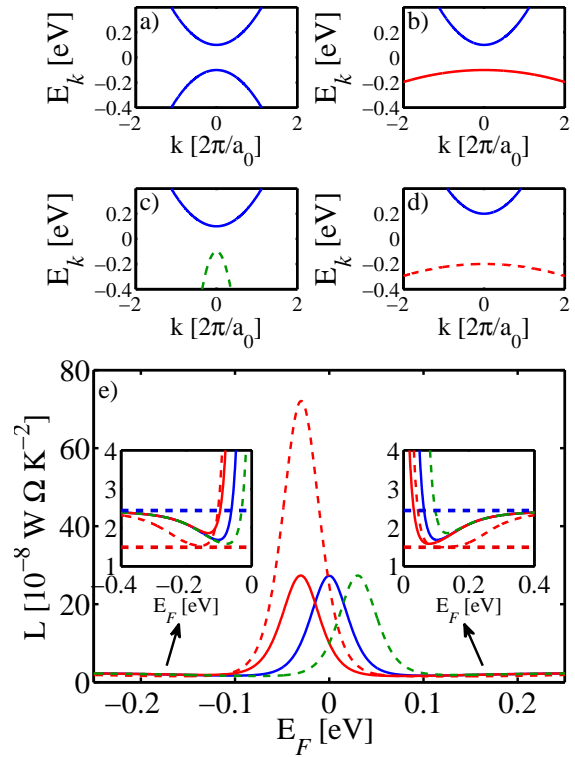}
\squeezeup
\caption{\label{Fig:Non_Interacting_1CB_1VB} Lorenz number versus Fermi level for a two-band bipolar system in the absence of inter-band scattering; one valence and one conduction band. The arrangement of this figure is identical to that of Figure~\ref{Fig:Non_Interacting_2CB}. The effective mass of the conduction band in all cases is $m_0$ and the effective mass of the valence band is taken to be $m_0$, $10m_0$, $0.1m_0$, $10m_0$ for panels a)-d) respectively. The band gap is 0.2 eV in all panels except d) where it is 0.4 (i.e. twice as large). Panel e) shows a very significant peak in the Lorenz number forming in the band-gap, which becomes large with increased band-gap and effective mass of the valence band. The insets show zoomed in regions, showing, by horizontal lines, both the degenerate (dashed blue) and non-degenerate (dashed red) limits.}
\end{figure}

\subsubsection{Two bands of opposite types: Lorenz number in bipolar systems}

There is an even larger deviation in the Lorenz number that occurs in bands with intra-band (but no inter-band) scattering in the case where one band is a conduction band and the other a valence when the band-gap is small (i.e. bipolar materials). Typical TE materials where this is true are BiTe, with a bandgap of $0.162$ eV, and PbTe, with a bandgap of $0.3$ eV at $T=300$ K.  \chgd{In this scenario, the fact that the Lorenz number deviates from the two single band limits is already well known.\cite{Kim15}  Therefore, in this section, we aim to derive a series of simple analytical expressions to estimate the size of deviation in this region based on simply bandstructure parameters and to demonstrate that the size of deviation can be quite substantial.} This increase in the mid-gap region is shown in Figure~\ref{Fig:Non_Interacting_1CB_1VB} in a similar panelled form to Fig.~\ref{Fig:Non_Interacting_2CB} with the valence band being the second, adjusted, band. Effective masses of the second band are the same as in that figure, except for Fig.~\ref{Fig:Non_Interacting_1CB_1VB}d which has a valence band effective mass $10 \times$ that of the conduction band (as in Fig.~\ref{Fig:Non_Interacting_1CB_1VB}b). The band-gaps in the figure are $0.2$ eV for all panels except d) where it is twice as large ($0.4$ eV).  The values of the band-gap were chosen to be simple and yet representative of those found in BiTe and PbTe. 

Looking at the Lorenz number plotted in Fig.~\ref{Fig:Non_Interacting_1CB_1VB}e, it is clear that there is an extremely drastic increase. A zoom-in can be seen in the insets and it is clear that the degenerate limit is recovered at the left (VB) and right (CB) sides.  \chgd{Saturation at the non-degenerate limit of course never occurs and, in fact, the real Lorenz value may be orders of magnitude larger.} Furthermore, it can be seen that the Fermi level of the peak depends on the relative effective masses of the valence and conduction bands, only being in the mid-gap in the case of equal masses (solid blue line). The heavier the effective mass of one band relative to the other, the closer the peak appears towards that band. Furthermore, looking at the case of the two curves of the same corresponding masses (the solid red and dashed red lines), but for different energy separations, it appears that the height of the peak is greatly affected by this separation. 

It is important to note that all curves are plotted such that the midgap is chosen to be $E_F=0$, and that the band profile described by the dashed red line actually has a different band-gap ($0.4$ eV instead of $0.2$ eV) than the other curves. Thus, although the peaks of the dashed and solid red lines (i.e. same effective masses, but different band-gaps) appear to coincide, suggesting that the peak location is not dependent on energy separation, this is merely a coincidence and we will discover the true relationship shortly.

The reason for this peak is intuitively simple.  The Seebeck coefficient is proportional to $I_1$ (see Eq.~\ref{Eq:I_integral_definition}), in which values at energies below the Fermi level contribute negatively and act to cancel the values at energies higher than it.  Thus, when the Fermi level is close to, or inside, a valence band, the sign of the Seebeck coefficient is opposite to when it is near or inside a conduction band.  As a result of this sign change, the Seebeck coefficient must be zero somewhere in the mid-gap.  As the Seebeck coefficient is \emph{subtracted} off the value of $\kappa_0$ (see Eq.~\ref{Eq:kappa_0_definition}), then $\kappa_e$ is enhanced in a situation where it becomes zero.  Thus, the Lorenz number is also enhanced.

As this peak lies in neither limit of the Fermi-Dirac integrals, it is difficult to completely model such behaviour with an analytical expression. However, given that this peak is strongly related to the Seebeck coefficient becoming zero one can estimate the Lorenz number at this zero point.  At this point it \emph{is} possible to develop an analytical expression. The task in doing so is two-fold, one must first determine for what value of the Fermi level the $S$ becomes zero and then determine the value of $L$ at that value of the Fermi level.

\begin{figure}[h]
\squeezeup \squeezeup
\includegraphics[width=0.45\textwidth]{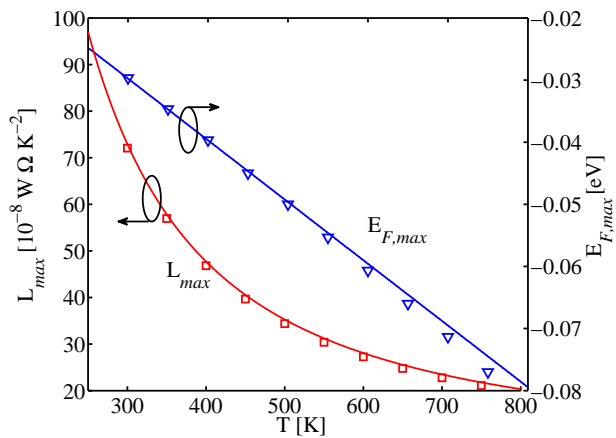}
\squeezeup \squeezeup
\caption{\label{Fig:L_max_analytical} Maximum Lorenz number of the midgap peak (left axis, red line and squares) and Fermi level location of said peak (right axis, blue line and triangles) as a function of temperature. In this plot the effective mass of the valence band is $10 \times$ that of the conduction (whose $m_*=m_0$) and the band-gap was $0.4$ eV. Symbols represent the results for full numerical calculation and lines represent the simplified expressions Eq.~\ref{Eq:L_Bipolar} and Eq.~\ref{Eq:Seebeck_Zero}.}
\end{figure}

The second task can be accomplished fairly straightforwardly by using the results of Eq.~\ref{Eq:kappa_0_both_bands} for the $\Delta$ value of: 
\begin{equation}
\label{Eq:Delta_BP}
\Delta_{BP} = -\frac{E_g}{k_B T} - 2 \eta_F,
\end{equation}
\noindent where $E_g$ is the band gap and the subscript $BP$ notes that this is the $\Delta$ for a \emph{bipolar} system. With this $\Delta_{BP}$ we have that:
\[ \eta_F + \Delta_{BP} = \frac{(E_C - E_g) - E_F}{k_B T} = \frac{E_V - E_F}{k_B T}. \] 
Substituting this into Eq.~\ref{Eq:kappa_0_both_bands} and dividing by temperature times Eq.~\ref{Eq:G_both_bands}, one can obtain the following expressions for the maximum Lorenz number:
\begin{widetext}
\begin{eqnarray}
\label{Eq:L_Bipolar}
L_{max} \approx L_0 + \left(\frac{k_B}{q} \right)^2 \left[ (r+2)^2 -  2 \eta_F^{max} (r+2)  + (\eta_F^{max})^2  + \frac{\Delta_{BP}}{1+\frac{\alpha_1}{\alpha_2} e^{-\Delta_{BP}}}  (2\eta_F^{max} - 2 (r+2) + \Delta_{BP}) \right],
\end{eqnarray}
\end{widetext}
\noindent where $L_0$ is the non-degenerate limit. It is important to note that this expression was derived under the assumption of the non-degenerate limit for the Fermi-Dirac integrals, such that $\mathcal{F}_r(\eta_F) \rightarrow \exp(\eta_F)$.  This approximation may seem questionable in a narrow band-gap system, but we will find in Fig~\ref{Fig:L_max_analytical} that its predictions are quantitatively accurate for the bandgap here of $0.2$ eV, which is comparable to that of the narrow band-gap semiconductors used in thermoelectrics (i.e. BiTe and PbTe).

The value of $\eta_F^{max}$ is that for which the Seebeck is zero. This must be determined numerically by finding the point where the expression in brackets in Eq.~\ref{Eq:S_both_bands} is zero (specifically, the zero root closest to the midgap):

\begin{equation}
\label{Eq:Seebeck_Zero}
0 = (r+2) - \eta_F^{max} - \frac{\Delta_{BP} }{\frac{\alpha_1}{\alpha_2} e^{-\Delta_{BP}} +1}.
\end{equation} 

The reason why $\eta_F^{max}$ must be solved numerically is because $\Delta_{BP}$ is also a function of $\eta_F$, as is shown in Eq.~\ref{Eq:Delta_BP}.

Using expression \ref{Eq:L_Bipolar}, evaluated at the point determined by solving Eq.~\ref{Eq:Seebeck_Zero}, it is possibly to obtain an estimate of the size of the Lorenz number deviation at its largest. Figure~\ref{Fig:L_max_analytical} shows a comparison of the maximum height (right $y$-axis, red line and squares) of the Lorenz number found in numerical calculation (squares) vs. the predictions of the analytical expression Eq.~\ref{Eq:L_Bipolar} (solid line) as a function of temperature. It also shows the Fermi level of the peak using both complete numerics (blue triangles) or by numerically solving the much simpler expression \ref{Eq:Seebeck_Zero} (blue line). It can be seen that the simpler expressions are still highly accurate. \chgd{The following expression can be considered valid for any value of $r$ and thus can be said to model not just acoustic phonons, but also strongly screened impurities ($r=0$), weakly screened ionized impurities ($r=2$) and optical phonons, provided the phonon energy is substantially smaller than the gap.}

\begin{figure}[h]
\squeezeup
\includegraphics[width=0.45\textwidth]{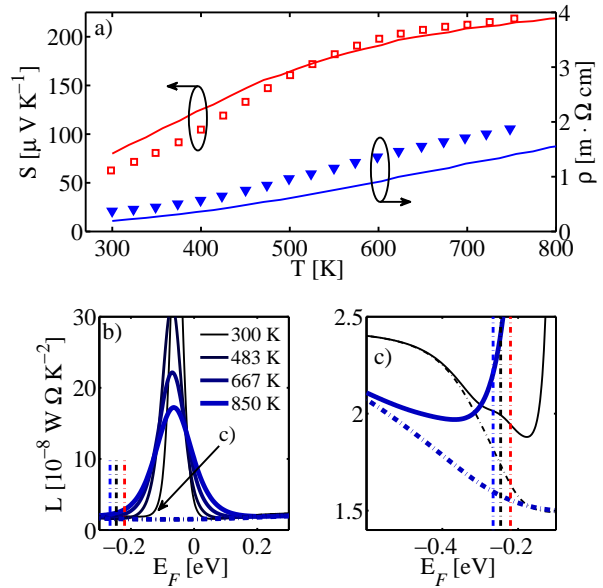}
\squeezeup
\caption{\label{Fig:PbTe_Calibration} Panel a) shows the Seebeck coefficient and resistivity vs. temperature in PbTe with comparison to experimental data extracted from Ref.~\onlinecite{Pei11}. Panel b) shows the Lorenz number versus Fermi level.The multiple lines represent a sample of curves at different temperature with the thinnest, darkest line being a temperature of $300 $K and the thickest bluest line being $850$ K. Dotted versions of those same lines, which show no peak, represent the result of calculating $L$ using Eq.\ref{Eq:Wrong_L_Formula}, \chgd{which ignores multi-band effects}. The Fermi levels indicated by dotted blue, black, and red vertical lines reflect positive carrier concentrations of $2.5 \times 10^{20}$ cm$^{-3}$, $2.0 \times 10^{20}$ cm$^{-3}$ and $1.5 \times 10^{20}$ cm$^{-3}$, respectively, and are used in Fig.~\ref{Fig:PbTe_Lorenz}. Panel c) is an enhanced plot of the left side of panel b).}
\end{figure}
\subsubsection{\label{PbTe_Section}Real Bipolar Material: The Case of PbTe}

As an example of the significant effect that multi-band Lorenz number deviations can have in real materials, even in the absence of inter-band scattering, we consider the case of the common thermoelectric material lead telluride (PbTe). This material has a fairly complex bandstructure which nicely encapsulates the effects that have already been discussed. In Ref.~\onlinecite{Pei11} it was shown that PbTe could be quantitatively matched to experimental data using a multi-band model with all inter-band scattering being ignored (i.e. only intra-band acoustic phonon scattering). In this section we will show that for a set of material parameters that describe the band-structure (adapted from Ref.~\onlinecite{Pei11}), the discrepancy between predictions made with Eq.~\ref{Eq:Wrong_L_Formula}, \chgd{which ignores multi-band effects}, and a more correct treatment can have a great effect on calculated Lorenz values and thus the error in the experimental determination of the lattice thermal conductivity.

Here we consider a simplified bandstructure of PbTe consisting of two valence bands ($\textrm{L}$ and $\Sigma$) and a single conduction band ($\textrm{C}$). We take the energy of the conduction band and the $\Sigma$ band to be fixed with the values of $E_{\textrm{C}} = 0.0$ eV and $E_C - E_{\Sigma} = E_{\textrm{C-}\Sigma }= 0.36$ eV. Conversely, we assume that the $\textrm{L}$ band energy changes with temperature according to the function (as used in Ref.~\onlinecite{Pei11}):
\begin{equation}
E_{\textrm{C-L}} = 0.09+\frac{4T}{10000}\;\; \mathrm{eV.}
\end{equation}
Thus, for low temperatures, the $\textrm{L}$ band is the highest energy valence band but at high temperatures there is a cross-over and for $T \gtrsim 450 $K, the $\Sigma$ band becomes higher in energy. The effective mass of the various bands are taken to be temperature dependent with the form
\begin{equation}
m^*(1 +\Delta m^*\frac{T-300 \textrm{K}}{T})
\end{equation}
where  $m^*_L=0.36m_0$, $m^*_{\Sigma}=2m_0$ and $m^*_C=0.3m_0$ and $\Delta m^*_{L} =0.5$, $\Delta m^*_{\Sigma} =0.0$ (i.e. no change) and $\Delta m^*_{C} =0.5$ (as in Ref.~\onlinecite{Pei11}). Inter-valley scattering is ignored and only intra-valley acoustic phonon scattering is considered. The strength of this is most easily calculated using the expression for the scattering time from Boltzmann transport theory:\cite{LundstromCarrierBook}
\begin{equation}
\label{Eq:tau_AP}
\tau_{AP} = \frac{\hbar \rho c_s^2}{\pi D_{AP}^2}
\end{equation}
where $\rho$ is taken to be $8.164$ g/cm$^3$, $D_{AP}$ is taken to be $19$ eV for the $\textrm{L}$ and $\textrm{C}$ bands and $9.5$ eV for the $\Sigma$ band and $c_{s}$ is taken to be $3600$ m/s (as in Ref.~\onlinecite{Pei11}). After calculating $\tau_{AP}$, the value is converted into a mean-free-path for back-scattering using Eq.~\ref{Eq:MFP_wrt_tau}, and the Landauer formalism is the approach used for final calculation.

All the values used here were adapted from Ref.~\onlinecite{Pei11}, however, it is important to note that in that work, the $\textrm{L}$ and $\textrm{C}$ bands were also assumed to be nonparabolic Kane bands. Although such an assumption would surely improve matches to experimental data, it is also an unnecessary complexity and a divergence from the analytical expressions and simple effective mass discussion considered in this work. Thus, all bands are treated as parabolic in our model. However, Fig.~\ref{Fig:PbTe_Calibration}a shows experimental results for the Seebeck coefficient and resistivity taken from experiments in Ref.~\onlinecite{Pei11} on undoped polycrystalline PbTe (squares). It is clear that this assumption of parabolicity (lines) still produces results that are accurate enough to motivate a discussion of the Lorenz number in real systems.

Fig.~\ref{Fig:PbTe_Calibration}b shows the Lorenz number versus Fermi level for the temperatures $300$ K to $850$ K (lower temperatures being thinner, blacker lines; higher temperatures being thicker, bluer lines). It also shows the result of using the \chgd{more accurate approach towards} calculating the Lorenz number (solid curves), which results in a substantial peak, and, as a reference, by using Eq.~\ref{Eq:Wrong_L_Formula} (dotted lines of the same color). The difference between the two methods can be seen more clearly in the zoomed-in Fig.~\ref{Fig:PbTe_Calibration}c. In that figure it is found that both methods agree in the degenerate limit (i.e. to the left), but in the middle of the band-gap the method of Eq.~\ref{Eq:Wrong_L_Formula}, which ignores coupling terms such as those proportional to $S_1 S_2$, shows no peak at all. The vertical blue, black and red lines indicate carrier concentrations of $2.5 \times 10^{20}$ cm$^{-3}$, $2.0 \times 10^{20}$ cm$^{-3}$ and $1.5 \times 10^{20}$ cm$^{-3}$, respectively, \chgd{corresponding to typical doping values for this material found in TE applications}. These points are shown here as they are used later on in Fig.~\ref{Fig:PbTe_Lorenz} below. 

Looking at Fig.~\ref{Fig:PbTe_Calibration}b-c it is clear that the effect of the peak in the Lorenz number in the mid-gap region is substantial at all temperatures, with the height of the peak being less at higher temperatures but its width being greater. Note that, the Eq.~\ref{Eq:Wrong_L_Formula} curves suggest that the entire mid-gap region has a Lorenz number corresponding to the non-degenerate limit, when in reality the Lorenz number differs from this value by orders of magnitude throughout the entire range of Fermi energy levels that reside in the bandgap.

\begin{figure}
\squeezeup \squeezeup
\includegraphics[width=0.45\textwidth]{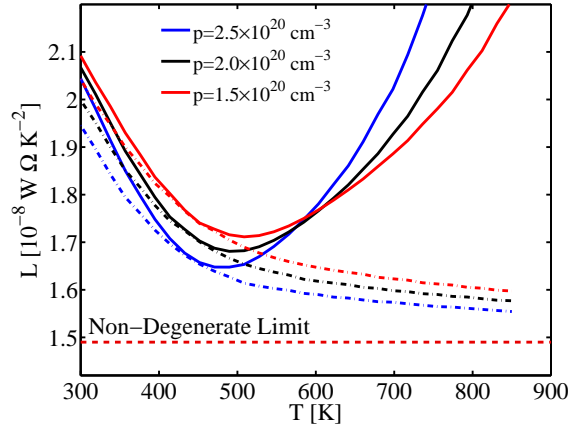}
\squeezeup \squeezeup
\caption{\label{Fig:PbTe_Lorenz} Lorenz number vs. temperature for different carrier concentrations for a two-band bipolar system. The dotted lines represent the Lorenz number as calculated using Eq.~\ref{Eq:Wrong_L_Formula}, where the solid lines represent calculations which correctly include cross-terms that occur even if there is no explicit coupling between bands.}
\end{figure}

Figure~\ref{Fig:PbTe_Lorenz} shows the Lorenz numbers as a function of temperature for the three carrier concentrations marked in Fig.~\ref{Fig:PbTe_Calibration}b-c. These carrier concentrations are found to be a reasonable range for $p$-doped PbTe.\cite{Pei11} The dotted lines represent the Lorenz number as calculated using Eq.~\ref{Eq:Wrong_L_Formula}, where the solid lines represent its value when the inter-band cross-terms, which appear even in bands not connected by inter-band scattering, are taken into account. Although the discrepancy is small at room temperature,\chgd{ and Eq.~\ref{Eq:Wrong_L_Formula} seems to provide an accurate estimate of the Lorenz number}, the difference is as large as a factor of 80\% for the case of $p=2.5\times 10^{20}$  at $T=850$ K (the maximum of the blue line, occurring outside the figure range, is $L \sim 2.9 \times 10^{-8} $ W $\Omega$ K$^{-2}$)).  \chgd{It is noted that PbTe is a TE material which is optimal for high temperature applications.  Thus, care needs to be taken when extracting the lattice thermal conductivity using the Lorenz number at these temperatures.}

As was previously discussed, these results suggest that the Lorenz number can be substantially under-estimated in a material like PbTe. As a result, since the lattice thermal conductivity is often extracted from experimental measurements of $\kappa_{tot}$ and the assumption of a Lorenz number in the non-degenerate or degenerate limit, the recorded values of $\kappa_{L}$ would correspondingly be over-estimated. As a result this may mean that nano-structuring attempts to lower lattice thermal conductivity may indeed by more successful than is recorded, but also that introducing further phonon scattering mechanisms to further reduce $\kappa_L$ may not result in lower thermal conductivities as $\kappa_L$ may already reach at,or below the amorphous limit, for example.
\begin{figure}[t]
\squeezeup
\includegraphics[width=0.42\textwidth]{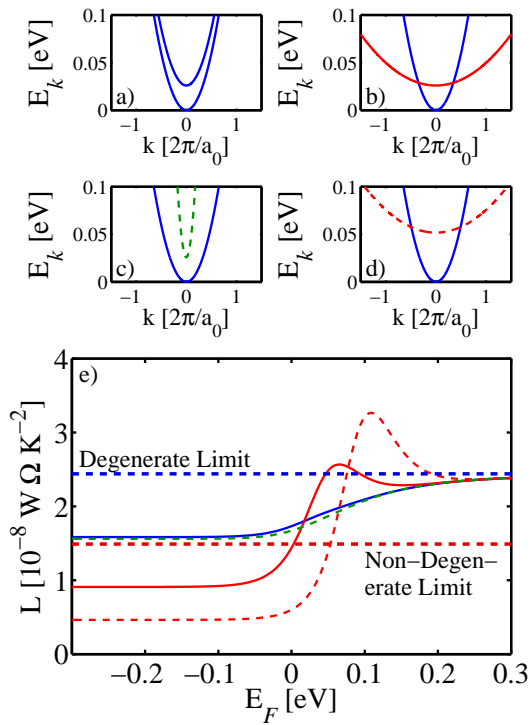}
\squeezeup
\caption{\label{Fig:Interacting_2CB} Lorenz number versus Fermi level for two conduction bands \emph{interacting} through inter-band scattering at $T=300$ K. The arrangement of this figure is identical to that of Figures \ref{Fig:Non_Interacting_2CB} and \ref{Fig:Non_Interacting_1CB_1VB}. The effective mass of the lower conduction band in all cases is $m_0$ and the effective mass of the upper conduction band is taken to be $m_0$, $10m_0$, $0.1m_0$, $10m_0$ for panels a)-d) respectively. The band separation is $k_B T$ in all panels except d) where it is $2 k_B T$ (i.e. twice as large). Panel e) shows clear non-monotonic behaviour in the Lorenz number in the intermediate regions between the degenerate (dashed blue) and non-degenerate (dashed red) limits. In addition, strong deviations from the expected non-degenerate limit can be seen for large band separation energies and effective masses}
\end{figure}

\subsection{\label{Interacting_Bands}Bands Interacting Through Inter-band Scattering}

In addition to the deviations discussed for the case of non-interacting bands (i.e. where only intra-band scattering is allowed), there are also additional Lorenz number deviations that occur once the multiple bands in a system are coupled by inter-band scattering. A panel figure similar to that of Fig.~\ref{Fig:Non_Interacting_2CB} is shown in Fig.~\ref{Fig:Interacting_2CB} for the case of a two conduction band system within the presence of inter-band scattering (at $T= 300$ K). For the sake of simplicity, the strength of inter-band scattering was taken to be the same as that for acoustic phonon scattering (i.e. $\lambda_{IB}=\lambda_{AP}$) and the final effective mean-free-path for back-scattering is given by the alternate Matthiessen's rule \chgd{(in terms of mean-free-paths rather than relaxation times):}
\begin{equation}
\frac{1}{ \lambda_{tot}(E)} = \frac{1}{ \lambda_{AP}(E)} + \frac{1}{ \lambda_{IB} (E)},
\end{equation}
\noindent which follows straightforwardly from the regular Matthiessen's rule in terms of scattering times and Eq.~\ref{Eq:MFP_wrt_tau}.

The key features of Fig.~\ref{Fig:Interacting_2CB} are: i) the deviation from expectation in the non-degenerate limit, and ii) the non-monotonic peak at intermediate Fermi levels between the two limits. In both cases, the amount of deviation increases with both the effective mass of the upper band relative to the lower band and the size of the energy separation between the two conduction bands, with the largest deviations being found for the case of Fig.~\ref{Fig:Interacting_2CB}d with an upper band effective mass of $10m_0$ and an energy separation of $2 k_B T$. 

\subsubsection{Reduction of $L$}

An extremely interesting feature of the Lorenz deviations in this case is that $L$ can, in fact, be less than expectation by as much as a factor of 3 (Fig.~\ref{Fig:Interacting_2CB}d) and thus this is the only case considered in this work where the $\kappa_L$ may be overestimated by a fair margin.  With respect to TE materials, one could imagine a case where $\kappa_L$ has not yet reached its amorphous limit value, and yet non-consideration of this type of Lorenz number deviation leads to the incorrect conclusion that no further lattice thermal conductivity reductions are possible.

The deviations in the non-degenerate limit are the result of the same effect discussed previously in Eqs.~\ref{Eq:Lorenz_2band_non_degenerate}.  Specifically, an ultimate result of the fact that $\kappa_e$ is more strongly weighted by higher energy contributions than $G$.  Thus, inter-band scattering into the upper band hinders $\kappa_e$ more than $G$ and thus decreases $L$. However, due to the scattering between the bands, the energy spectrum of the mean-free-path for back-scattering (or scattering time), no longer has a simple power-law form and thus a simple analytical result is not possible. Furthermore, looking at the size of the deviations from the non-degenerate limit it is clear that the effect is much larger when the bands scatter between one another and thus Eq.~\ref{Eq:Lorenz_2band_non_degenerate} is not valid and must only be considered as a special case of no or weak inter-band scattering.

The non-monotonic peak in the Lorenz number is another new feature that deserves further discussion.
\begin{figure}[t]
\squeezeup
\includegraphics[width=0.45\textwidth]{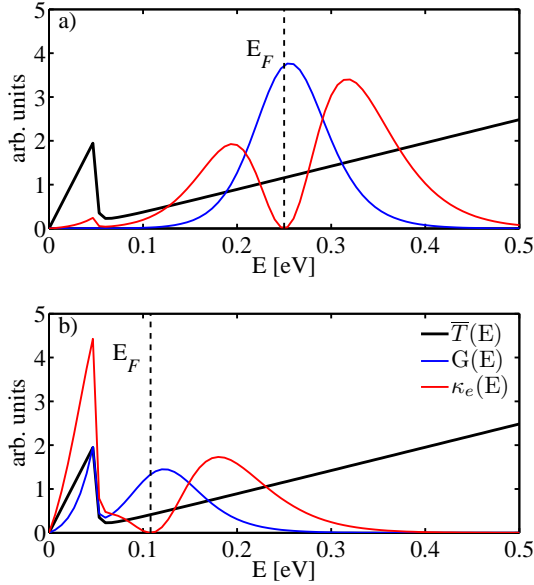}
\squeezeup
\caption{\label{Fig:Interacting_2CB_Transmission} The effective transmission ($\overline{T}(E) = T(E)M(E)$), energy resolved conductance ($G(E)$) and energy resolved electron thermal conductivity for two different Fermi levels: the first, shown in panel a), being a Fermi level well into the degenerate limit ($E_F = 0.25$ eV) and the second, shown in panel b), being the Fermi level value corresponding to the peak of the dashed red line in Fig.~\ref{Fig:Interacting_2CB}e ($E_F = 0.11$ eV). The effective mass and energy separation are those of panel e) of Fig.~\ref{Fig:Interacting_2CB} (i.e. $m_2=10m_1$, and a separation of $2 k_B T$). All quantities are plotted in arbitrary units chosen so that all curves can be clearly seen on the same graph, and qualitative comparisons made.} 
\end{figure}
\subsubsection{Non-monotonic behaviour}

Unlike Lorenz number deviations occurring in the non-degenerate limit, the non-monotonic behaviour seen in Fig.~\ref{Fig:Interacting_2CB}e,  is entirely attributable to the effect of inter-band scattering and does not occur in a system of independent bands. The key parameter in understanding this behaviour is the effective transmission given in Eq.~\ref{Eq:Effective_Transmission}. The source of the effect can ultimately be traced to two crucial facts: i) that the presence of inter-band scattering produces a sharp dip in the effective transmission function, and ii) that the electronic thermal conductance, dictated by $I_2$ (from Eq.~\ref{Eq:I_integral_definition}), differs from the electronic conductance, $G$, (which is $\propto I_0$) by a weighting factor of $\eta^2=(E-E_F/k_B T)^2$, which weights higher energies more. 

Figure \ref{Fig:Interacting_2CB_Transmission} shows the effective transmission ($\overline{T}(E) = T(E)M(E)$) (black line), energy resolved conductance ($G(E)$) (blue line) and energy resolved electronic thermal conductivity ($\kappa_e$) (red line) \chgd{for the Fermi levels of $E_F = 0.25$ eV and $E_F = 0.11$ eV. The first case shown is that where the peak forms in Fig.~\ref{Fig:Interacting_2CB}e, i.e. where the non-monotonic behaviour of the Lorenz number is most pronounced and} specifically where the upper band has an energy offset of $2k_B T$ and an effective mass that is $10 \times$ larger than that of the lower band. The sharp feature in the effective transmission at an energy of $2 k_B T$ is a result of inter-band scattering into the second band. Scattering is proportional to the density-of-states available to scatter into. Thus, when a second band, especially one with a very large effective mass (and thus large density-of-states), enters the energy window around the Fermi level, scattering increases vigorously and the mean-free-path for back-scattering (and thus $T(E)$) decreases.

This sharp feature in the effective transmission that results from inter-valley scattering is the reason for the non-monotonic behaviour in the Lorenz number. Fig.~\ref{Fig:Interacting_2CB_Transmission}a shows the behaviour of $\overline{T}(E)$ and $G(E)$ deep into the band ($E_F = 0.25$ eV), where the Lorenz number is saturated at its degenerate limit. The conductance, $G(E)$, is peaked near the Fermi level, whereas the electronic thermal conductance, $\kappa_e(E)$, has two peaks lying some distance to either side. Conversely, Fig.~\ref{Fig:Interacting_2CB_Transmission}b shows a Fermi level of $0.11$ eV, which corresponds to the peak of the non-monotonic behaviour in Figure \ref{Fig:Interacting_2CB}e. Looking at the figure, one can clearly see the reason for the enhancement - the conductance peaks at the pronounced dip in the effective transmission, whereas $\kappa_e$ straddles this dip, with the left-most (i.e. lower in energy) of its two peaks lying right at the top of the sharp transmission feature.  Thus $\kappa_e/G$ is enhanced.

Thus, the degree of non-monotonicity in interacting multi-band systems is driven by the sharpness of the jagged feature in the effective transmission. The sharpness of this feature is, in turn, dependent on how much greater the density-of-states of the upper band is and how easy it is for carriers to scatter from one band to the other. Therefore, as the inter-band scattering amplitude, the ratio of effective masses (i.e. $m^*_2/m^*_1$) and the size of the energy separation between the bands increases, this non-monotonic deviation of the Lorenz number becomes a greater effect. This is also why such non-monotonic features do not appear in band systems without inter-band scattering, as they are dependent on the sharp dip in the the effective transmission, which only occurs when this scattering is considered.

\subsection{\label{SnSe_Section} Multi-bands - The example of SnSe}

As a final synthesis of all aspects considered in this work, we consider a three-band model - two valence bands plus one conduction band - of the promising new thermoelectric material, $p$-type tin selenide (SnSe).\cite{Zhao14,Zhao16}  In this three-band model we include inter-band scattering between the two valence bands and, thus, the model includes all three aspects previously discussed (i.e. mid-gap deviations, as well as scattering and non-scattering driven two-band deviations). Despite SnSe's complex band-structure, we will show that this simpler three-band model can adequately match experimental results for $p$-doped SnSe, and therefore, can be used to explore the \chgd{Lorenz number values in a more detailed and accurate manner.}  

\chgd{It is worth pointing out that the assumption that PbTe has no inter-band scattering but SnSe does, as well as the assumptions of parabolic bands, are justified here only by the simple fact that these assumptions match experimental data adequately.}  Undoubtedly, better matches to experimental data could be achieved by including more scattering mechanisms \chgd{(such as optical phonons)}, more bands, non-parabolic effects, etc.  However, it is not the goal of this work to achieve the most quantitatively accurate possible model of these materials but merely to highlight the approximate size of deviations that can result from the effects discussed here.

\begin{figure}[t]
\squeezeup \squeezeup
\includegraphics[width=0.51\textwidth]{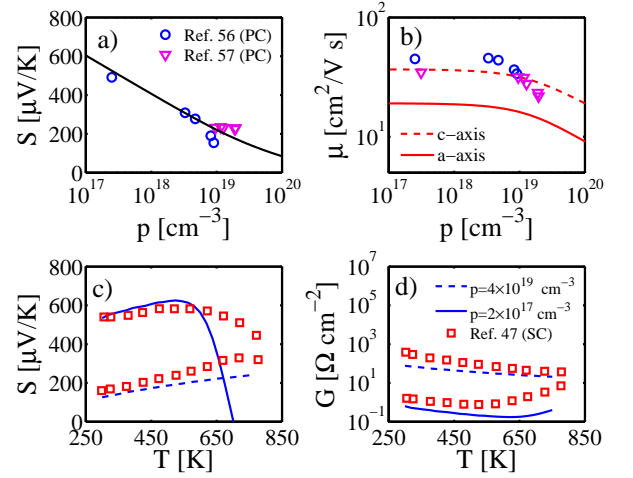}
\squeezeup \squeezeup \squeezeup
\caption{\label{Fig:SnSe_Calibration} Calibration figures demonstrating the ability of the three-band model to match experimental data for SnSe. Panel a) and b) show the $a$-axis Seebeck coefficient (solid black) and mobility (solid red) respectively as a function of carrier concentration, with the blue dots and pink triangles representing experimental data on polycrystalline (PC) samples taken from Ref.~\onlinecite{Chen14} and Ref.~\onlinecite{Leng16} respectively.  The dotted red line in b) represents mobility along the $c$-axis. Panels c) and d) show the Seebeck coefficient and conductance, respectively, as a function of temperature for a lightly (solid blue) and heavily (dashed blue) doped single crystalline (SC) sample.  Red squares represent experimental results from Ref.~\onlinecite{Zhao16} of a SC sample along the $a$-axis. }
\end{figure}

We take as a model of SnSe a system of two valence bands, with inter-band acoustic phonon scattering (i.e. $r=0$ in Eq.~\ref{Eq:Lambda_temp_form}) and a conduction band separated by a band-gap (i.e. a three-band model).  Scattering is included in a manner identical to that in Sec.~\ref{PbTe_Section} for PbTe, with the only exception being that inter-band acoustic phonon scattering is included, and its strength is taken to be the same as intra-band scattering (i.e. $ \lambda_{AP} = \lambda_{IB}$).  The values for filling Eq.~\ref{Eq:tau_AP} were drawn from the Density Functional Theory (DFT) results of Ref.~\onlinecite{Guo15} for the $a$-axis, which was found to match most closely the experimental data for single crystalline samples and the $c$-axis, which was found to match the polycrystalline mobility data (see Fig.~\ref{Fig:SnSe_Calibration}b).  The mass density, $\rho$, was taken to be $6.179$ g/cm$^3$, the sound velocity, $c_s$, was taken to be $3356$ m/s for the $a$-axis ($3267$ m/s along the $c$-axis), the deformation potentials of the valence and conductions bands were taken to be $14.1$ eV ($15.8$ eV for the $c$-axis) and $12.9$ eV ($13.2$ eV for the $c$-axis) respectively.  A value of the band-gap of $0.78$ eV was also assumed based on that work as well as effective mass values for the conduction band of $m_{x}^* = 0.5m_0$, $m_{y}^* = 0.12m_0$ $m_{z}^*=0.16m_0$.

The values of the valence band effective masses were taken from the more recent Ref.~\onlinecite{Zhao16} where they were found to match experimental data for degenerately doped single crystalline samples.  For the first (i.e. highest energy) valence band, $m_{x}^* = 0.76 m_0$, $m_{y}^* = 0.33 m_0$, and $m_{z}^* = 0.14 m_0$. For the second valence band, the effective masses  $m_{x}^* = 2.49 m_0$, $m_{y}^* = 0.18 m_0$, and $m_{z}^* = 0.19 m_0$ are heavier than for the first band.  All bands are doubly degenerate and a band-separation of $0.06$ eV was also assumed based on that work.

It is important to re-iterate how effective mass enters into the Landauer formalism (through the density-of-modes) versus how it is calculated in the density-of-states, which is necessary to determine carrier concentrations.\cite{Lundstrom_Book13} Along the $a$-axis, the \emph{density-of-modes} effective mass is taken here to be $g_b \sqrt{m_{y}^* m_{z}^*}$, where $g_b$ is the degeneracy of the band. This, in essence, represents a cross-section of the effective mass in the plane perpendicular to the transport direction. Conversely, here the \emph{density-of-states} effective mass is given by $g_b^{2/3}(m_{x}^* m_{y}^* m_{z}^*)^{1/3}$.

With this distinction in mind, a three-band model of SnSe is used in Fig.~\ref{Fig:SnSe_Calibration} to compare against experimental results. Fig.~\ref{Fig:SnSe_Calibration}a and Fig.~\ref{Fig:SnSe_Calibration}b represent the  Seebeck coefficient and mobility respectively as a function of carrier concentration.  Carrier concentration is calculated by integrating the density-of-states (using the density-of-states effective mass) and the mobility is calculated from $\mu = G/q p$, where $p$ is the positive carrier concentration.  The solid curves - black for the Seebeck coefficient, red for the mobility - represent transport along the $a$-axis and the dotted red curve represents the mobility along the $c$-axis.  The blue circle and purple triangles represent experimental data on silver (Ag) doped polycrystalline (PC) SnSe taken from Ref.~\onlinecite{Chen14} and Ref.~\onlinecite{Leng16}.  As that data is taken from a polycrystalline sample, and thus each grain has a different orientation, both the $a$- and $c$-axis calculations are shown. It is clear that this relatively simple model adequately matches the data in both the undoped regime($p \approx 2 \times 10^{17}$ cm$^{-3}$) and the highly doped regime ($p \approx 10^{19}$ cm$^{-3}$).

As a second demonstration of the quantitative validity of the model, it is compared against experimental results from Ref.~\onlinecite{Zhao16} for single crystalline (SC) SnSe along the $a$-axis.  Doping was done with sodium (Na).  Fig.~\ref{Fig:SnSe_Calibration}c and Fig.~\ref{Fig:SnSe_Calibration}d show the Seebeck coefficient and conductance, respectively, versus temperature for the case of no doping (indicated in Ref.~\onlinecite{Zhao16} to occur at $p\approx 2\times10^{17}$, shown by the dashed blue lines in  both figures), and the highly-doped case (indicated in Ref.~\onlinecite{Zhao16} to occur at $p \approx 4\times10^{19}$, shown by the solid blue lines in  both figures).  It is important to note that SnSe undergoes a structural phase-transition at $\approx 750$ K, which is in no way captured by the simple model here.  

The existence of this structural phase-transition, however, does not likely explain all of the deviations in the Seebeck coefficient from experiment observed at high temperatures \chgd{in the lightly-doped case}.  The cause of this deviation is the fact that the calculated Seebeck data (solid blue line) dramatically tends towards zero at temperatures higher than approximately $600$ K, where the experimental data have a more gradual downturn.  At higher temperatures, in order to preserve a fixed carrier concentration, the Fermi level must move away from the band and closer to the mid-gap.  It is this approach towards the mid-gap that causes the Seebeck coefficient to tend towards zero.  This drop towards zero can be made to occur at higher temperatures by either considering a larger band-gap, or by increasing the average effective mass of the conduction bands (thus pulling the zero Seebeck point closer to the conduction band as described by Eq.~\ref{Eq:Seebeck_Zero} and illustrated in Fig.~\ref{Fig:Non_Interacting_1CB_1VB}).  This is to say, that it is possible to obtain a closer fit to experiment in this high-temperature range by adjusting the band-gap and band effective masses of the model. However, the parameters used here were extracted from either experiment or density functional theory calculations and, thus, changing their value in an ad hoc manner is difficult to justify and, given that the goal of this work is not to produce a maximally accurate model, is of little value.  This is in addition to the expected inaccuracy of the model regardless, due to the structural phase-transition, at high temperatures.  With this understanding, the quantitative fit to data is adequate for the purposes of exploring Lorenz number behaviour in this material, at least for temperatures below $\sim 600$ K, \chgd{especially for the highly-doped case, which is more relevant for TE applications.}

\begin{figure}[t]
\squeezeup \squeezeup \squeezeup
\includegraphics[width=0.45\textwidth]{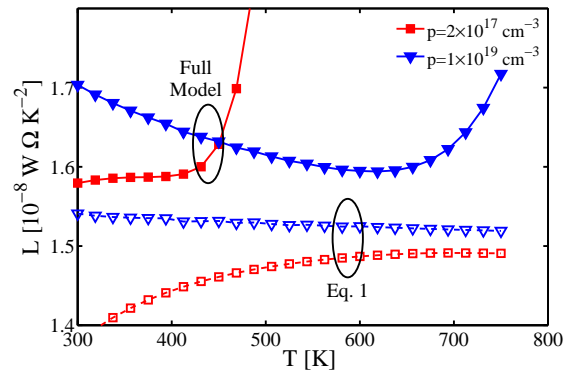}
\squeezeup \squeezeup \squeezeup
\caption{\label{Fig:SnSe_Figure} Lorenz number versus temperature in tin selenide (SnSe).  The red squares indicate a $p$-doping level of $2\times10^{17}$ cm$^{-3}$ reflecting an essentially undoped sample, where blue triangles represent a doping of $10^{19}$ cm$^{-3}$ reflecting heavy doping (at $p\approx 4\times10^{19}$ cm$^{-3}$ the Fermi level enters the valence band).  Dotted lines represent predictions based on Eq.~\ref{Eq:Wrong_L_Formula} and solid lines represent results from the complete inclusion of multi-band effects.}
\end{figure}

Since the three-band model has been shown to acceptably reproduce experimental results for the Seebeck coefficient and conductance, it is worthwhile to consider what it says about the Lorenz number.  Fig.~\ref{Fig:SnSe_Figure} shows the calculated Lorenz number versus temperature at two different carrier concentrations: $p=2\times 10^{17}$ cm$^{-3}$ in red squares, representing undoped SnSe with a Fermi level in the mid-gap region, and $p=1\times 10^{19}$ cm$^{-3}$, in blue triangles, representing a heavily doped sample (in Ref.~\onlinecite{Zhao16} it was estimated that $p\approx 4 \times 10^{19}$ cm$^{-3}$ corresponds to the point where the Fermi level enters the valence band).  The solid curves with solid markers represent values from the Landauer approach taken here, and the dotted curves with hollow markers represent those resulting from a Lorenz number calculation based on Eq.~\ref{Eq:Wrong_L_Formula}.  In the highly doped case (blue lines) the effect is noticeable, though arguably not substantial, amounting to a $\sim 15\%$ under-estimation of the Lorenz number at low and high temperatures.  Though it is important to re-iterate that at high-temperatures, SnSe actually undergoes a structural phase-transition, and, thus, this model may not quantitatively reflect that material in the high temperature range, \chgd{although in the highly-doped case, which is more relevant for TE applications, a greater agreement is achieved (this is seen by comparing the blue lines).} However, in the undoped case the effect is undoubtedly significant.  For low temperatures, the discrepancy is, again, $\sim 15\%$.  However, for intermediate and high temperatures the deviation from expectation differing by several orders of magnitude (red solid lines goes off the graph at very high values).

\section{Conclusions}

In this work, we have explored the types and sizes of deviations of the Lorenz number from the expected non-degenerate limit of $L_0= 2 (k_B/q)^2 =1.49\times 10^{-8}$ W $\Omega$ K$^{-2}$ and the expected degenerate, or metallic, limit of $L_0=\pi^2/3 (k_B/q)^2=2.45 \times 10^{-8}$ W $\Omega$ K$^{-2}$ that occur due to multi-band effects.  Specifically, we have shown that the Lorenz number can deviate markedly from expectation in the case of multiple-bands of the same type (i.e. multiple conduction bands or multiple valence bands), even if there is no explicit inter-band scattering present.

For the deviations outlined in this work, a number of analytical expressions were derived that allow for \chgd{a more accurate estimation of} the Lorenz number.  Furthermore, particular stress is placed on the amount of error that can result from the assumption that multi-band systems can be treated as entirely decoupled, \chgd{resulting in an equations like} Eq.~\ref{Eq:Wrong_L_Formula}.  In this case it was found that values can deviate by orders of magnitude in the mid-gap of bipolar systems and by approximately a factor of two for unipolar systems. \chgd{The study was performed within the Boltzmann transport approach expressed within the Landauer form with an assumed semi-classical transmission function.  Thus, it can be said to be generally relevant to diffusive transport for any scattering mechanism that follows a power-law form in energy.}

\chgd{The primary model used in this work is that of parabolic bands in the presence of acoustic scattering only.  For materials where such a model is inappropriate, it is likely that the main insights obtained can be said to be only qualitatively accurate, although providing even qualitative insights on how multiple bands and intra- vs inter-valley scattering affect the Lorenz number is of great value.}  However, as a demonstration of the real effect these deviations can have on real materials, specific examples of the common thermoelectric materials lead telluride (PbTe) and tin selenide (SnSe) were explored.  Simple multi-band models were developed for each material and found to adequately match published experimental results.  From these models, the value of the Lorenz number was explored as a function of temperature and doping (or Fermi level).  It was found that substantial deviation from expectation can occur in some cases in these materials.

These deviations have important consequences to experimental results, as the Lorenz number is often used as a means of calculating the lattice thermal conductivity from measurements of the total thermal conductivity.  Thus, deviations in the Lorenz number represent a mis-estimation of the relative contribution of the lattice, versus charge carriers, to the total thermal conductivity.  This is of particular importance in the field of thermoelectrics, where there is a strong push to lower the lattice thermal conductivity and there is constant debate as to whether lattice thermal conductivity values have reached theoretical amorphous limits.

\section{Acknowledgements}

M.T. has been supported by the Austrian Research Promotion Agency (FFG) Project No. 850743 QTSMoS. N.N. has received funding from the European Research Council (ERC) under the European Union’s Horizon 2020 Research and Innovation programme (Grant Agreement No. 678763).

\appendix*
\section{\label{Appendix1}Derivation of Transport Coefficients in a Two-Band System in the Absence of Inter-band Scattering}

The values of $\overline{SG}$ and $\kappa_0$ for the full system of two conduction or valence bands without inter-band scattering can be obtained through addition. For a 3D parabolic band in the non-degenerate limit, these quantities take the form:
\begin{eqnarray}
\overline{SG}_b &\approx &  -  \frac{k_B}{q} \gamma \Gamma(r+2) \alpha_b e^{\eta_{F_b}} \left((r+2) - \eta_{F_b}  \right)  \\
\kappa_{0,b} &\approx & T \left(\frac{k_B}{q} \right)^2 \gamma \Gamma(r+2) \alpha_b e^{\eta_{F_b}}  \\* \nonumber &\times &\left( (r+3)(r+2) - 2\eta_{F_b}(r+2)  + \eta_{F_b}^2 \right),
\end{eqnarray}
\noindent and adding them for both bands produces the expressions:
\begin{widetext}
\begin{eqnarray}
\label{Eq:SG_both_bands}
\overline{SG} \approx -  \frac{k_B}{q} \gamma \Gamma(r+2) e^{\eta_F} \psi \left( (r+2) - \eta_F - \frac{\Delta \alpha_2 e^{\Delta}}{\psi} \right) \\
\label{Eq:kappa_0_both_bands}
\kappa_0 \approx T \left(\frac{k_B}{q} \right)^2 \gamma \Gamma(r+2) e^{\eta_F} \psi \left[ (r+2)^2 +(r+2) -2 \eta_F (r+2)   + \eta_F^2  + (2\eta_F - 2 (r+2) + \Delta) \frac{\Delta \alpha_2 e^{\Delta}}{\psi} \right].
\end{eqnarray}
\end{widetext}
It is important to re-iterate that although $\overline{SG}$ and $\kappa_0$ add for independent bands, $\kappa_e$ does not (see Eq.~\ref{Eq:kappa_e_definition} where the $S^2$ term couples bands). Using Eqs.~\ref{Eq:G_both_bands}, \ref{Eq:SG_both_bands} and \ref{Eq:kappa_0_both_bands}, and inserting them into Eqs.~\ref{Eq:S_definition} and \ref{Eq:kappa_e_definition} we obtain the expressions Eq.~\ref{Eq:S_both_bands} and Eq.~\ref{Eq:kappa_both_bands}


\bibliography{PRB_2016_Lorenz2}

\end{document}